\documentclass[twocolumn,prl,aps,superscriptaddress,floatfix]{revtex4-1}
\usepackage{graphicx}
\usepackage{amsmath}
\usepackage{natbib}
\usepackage{gensymb}

\begin{document}

\title{Engineering the strain and interlayer excitons of 2D materials via lithographically engraved hexagonal boron nitride}

\author{Yu-Chiang Hsieh}
\affiliation{Department of Physics, National Cheng Kung University, Tainan 701, Taiwan}

\author{Zhen-You Lin}
\affiliation{Department of Physics, National Cheng Kung University, Tainan 701, Taiwan}

\author{Shin-Ji Fung}
\affiliation{Department of Physics, National Cheng Kung University, Tainan 701, Taiwan}

\author{Wen-Shin Lu}
\affiliation{Institute of Atomic and Molecular Sciences, Academia Sinica, Taipei 106, Taiwan}
\affiliation{Department of Electrophysics, National Yang Ming Chiao Tung University, Hsinchu 300, Taiwan}

\author{Sheng-Chin Ho}
\affiliation{Department of Physics, National Cheng Kung University, Tainan 701, Taiwan}

\author{Siang-Ping Hong}
\affiliation{Department of Physics, National Cheng Kung University, Tainan 701, Taiwan}

\author{Sheng-Zhu Ho}
\affiliation{Department of Physics, National Cheng Kung University, Tainan 701, Taiwan}

\author{Chiu-Hua Huang}
\affiliation{Department of Physics, National Cheng Kung University, Tainan 701, Taiwan}

\author{Kenji Watanabe}
\affiliation{Research Center for Functional Materials, National Institute for Materials Science, Namiki 1-1, Tsukuba, 305-0044, Ibaraki, Japan}

\author{Takashi Taniguchi}
\affiliation{International Center for Materials Nanoarchitectonics, National Institute for Materials Science, Namiki 1-1, Tsukuba, 305-0044, Ibaraki, Japan}

\author{Yang-Hao Chan}
\affiliation{Institute of Atomic and Molecular Sciences, Academia Sinica, Taipei 106, Taiwan}
\affiliation{Physics Division, National Center for Theoretical Sciences, Taipei 106, Taiwan}

\author{Yi-Chun Chen}
\email{ycchen93@ncku.edu.tw}
\affiliation{Department of Physics, National Cheng Kung University, Tainan 701, Taiwan}
\affiliation{Center for Quantum Frontiers of Research \& Technology (QFort), National Cheng Kung University, Tainan 701, Taiwan}

\author{Chung-Lin Wu}
\email{clwuphys@ncku.edu.tw}
\affiliation{Department of Physics, National Cheng Kung University, Tainan 701, Taiwan}
\affiliation{Center for Quantum Frontiers of Research \& Technology (QFort), National Cheng Kung University, Tainan 701, Taiwan}

\author{Tse-Ming Chen}
\email{tmchen@phys.ncku.edu.tw}
\affiliation{Department of Physics, National Cheng Kung University, Tainan 701, Taiwan}
\affiliation{Center for Quantum Frontiers of Research \& Technology (QFort), National Cheng Kung University, Tainan 701, Taiwan}

\begin{abstract}
Strain engineering has quickly emerged as a viable option to modify the electronic, optical and magnetic properties of 2D materials. However, it remains challenging to arbitrarily control the strain. Here we show that by creating atomically-flat surface nanostructures in hexagonal boron nitride, we achieve an arbitrary on-chip control of both the strain distribution and magnitude on high-quality molybdenum disulfide. The phonon and exciton emissions are shown to vary in accordance with our strain field designs, enabling us to write and draw any photoluminescence color image in a single chip. Moreover, our strain engineering offers a powerful means to significantly and controllably alter the strengths and energies of interlayer excitons at room temperature. This method can be easily extended to other material systems and offers a promise for functional excitonic devices.
\end{abstract}

\maketitle

The outstanding mechanical properties of two-dimensional (2D) materials such as the unprecedented stretchability not only create the opportunity for strain engineering but in doing so open a myriad of possibilities for future electronics and optoelectronics. In the bulk materials, strain engineering is conventionally only for improving the material and device quality, e.g., it is widely used to enhance the carrier mobility in semiconductor technology. However, strain engineering in 2D materials becomes extremely powerful and versatile due to the ability to substantially modify the band structure and completely change the materials' electronic and optoelectronic properties. For example, strain can open a band gap in graphene, turning its electronic state from a gapless semimetal to a gapped semiconductor\cite{Ni_2008}. It can also bring graphene into a topological insulator state with quantum valley Hall effect\cite{Guinea_2010}, a superconductor\cite{Si_2013}, or an exotic phase of matter with nonlinear Hall effect and valley-orbit coupling\cite{Ho_NE2021}, depending on the type of strain and exactly how it is induced. Beyond graphene, strain engineering can be universally used for all types of 2D materials and has extensively advanced fundamental science and technological applications in such diverse areas as electronics, optoelectronics, magnetism, spintronics / valleytronics, superconductivity, topological phases of matter, and energy harvesting\cite{Dai_AM2019}.

Although strain in 2D materials continues to capture the imagination of scientists and engineers alike, methods that are used for creating and controlling it are still limited. Many rely on the use of a flexible substrate to stretch or bend the 2D materials\cite{Ni_2008,Conley_NL2013,Castellanos_NL2013,Rice_PRB13}, thereby introducing a global, uniform strain (lattice deformation). Others focus on nanoscale strain induced by surface nanostructures such as ripples and bubbles, which can either naturally (and hence in most cases randomly) occur in 2D materials\cite{Guinea_2010,Levy_S2010,Castellanos_NL2013,Li_PRL20} or be deliberately created by conforming 2D materials onto patterned substrates\cite{Ho_NE2021,Li_NC2015,Li_NM2016,Choi_NL2015,Hsu_SA20,Kang_NV2021}. Prior studies using such substrate engineering approaches have mostly relied on the use of epitaxial substrates with ripple, pyramid, nanocone arrays and so on \cite{Li_NC2015,Li_NM2016,Choi_NL2015,Hsu_SA20,Kang_NV2021}, and have recently been extended to lithographically defined nanostructures\cite{Ho_NE2021}. These have all brought significant and different impacts to strain engineering of 2D materials. However, up to now, an arbitrary control of the strain magnitude and distribution has not yet been demonstrated. More importantly, strain engineering via nanostructured substrates often suffers from poor interface quality with impurities and unwanted strains, which cause difficulties in investigating delicate quantum phenomena. All these challenges limit the extent to which strain engineering of 2D materials can effectively be used to advance both fundamental science and practical technologies.

\begin{figure*}
\centering 
\includegraphics[width=0.75\textwidth]{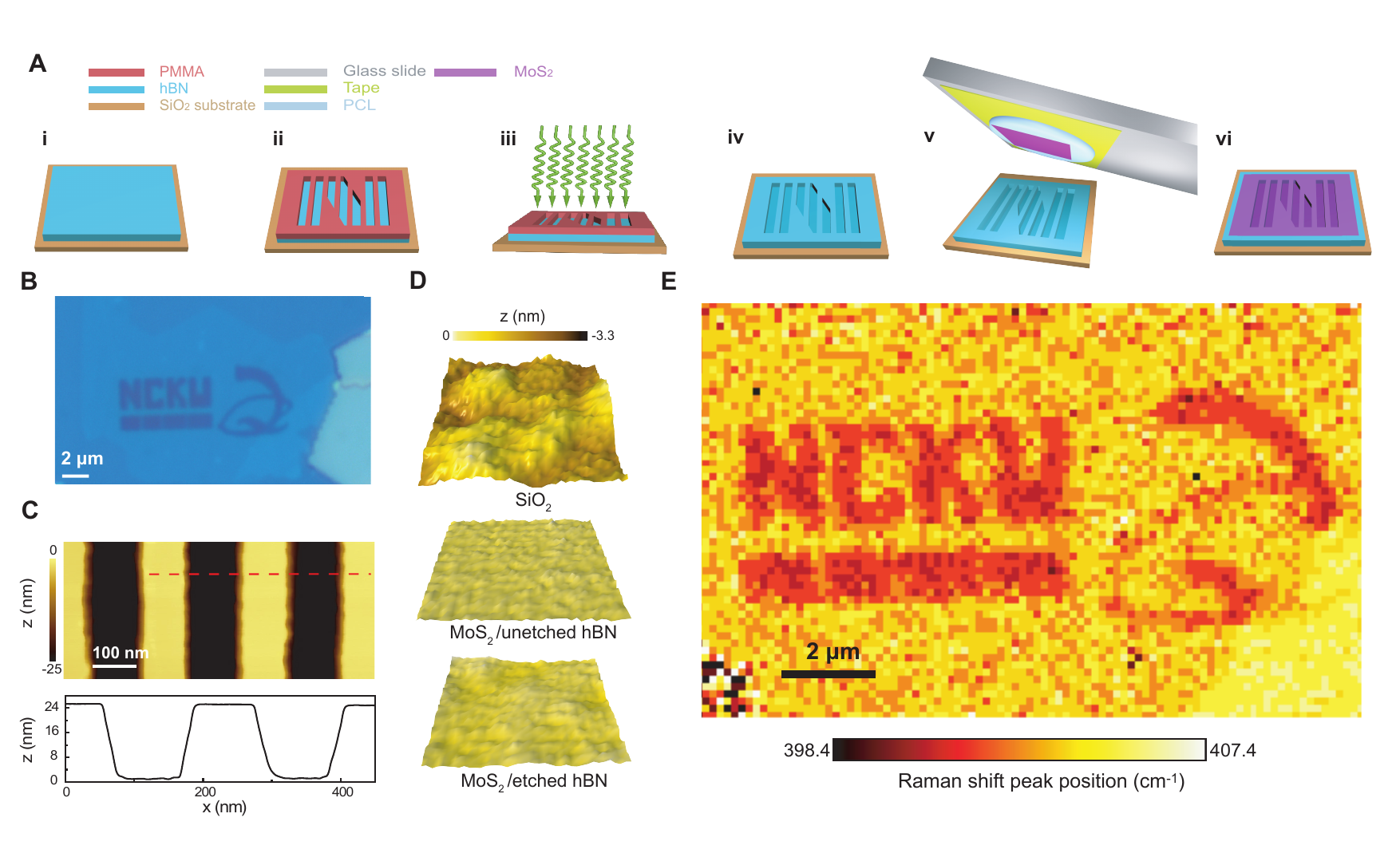}
\caption{\textbf{Arbitrary control of the local strain in high-quality 2D materials.} (\textbf{A}) Schematic illustrations from (\textbf{i}) to (\textbf{vi}) showing the fabrication procedure of the strain engineered MoS$_2$ devices. Selected hBN candidates are patterned by electron-beam lithography and etched by inductively coupled plasma (ICP) to obtain any desired topography on hBN. The target MoS$_2$ flake is then placed and conforms to the engraved hBN substrate to engineer the lattice strain in accordance to the substrate design. (\textbf{B}) Optical micrograph of the strain engineered MoS$_2$/hBN device. The engraved letters `NCKU' and QFort logo are composed of many 100-nm-wide trenches. (\textbf{C}) Atomic force microscopy image showing these 100-nm-wide trenches in the strain engineered MoS$_2$/hBN devices. The height profile shown in the bottom panel corresponds to the red dashed line in the AFM image.  (\textbf{D}) Comparison of surface roughness for a standard SiO$_2$ substrate (0.35~nm), the MoS$_2$ on the unetched hBN region (0.08~nm), and the MoS$_2$ on the etched hBN region (0.12~nm) in our devices. (\textbf{E}) Spatial map of the Raman spectroscopy for the A$_{1g}$ phonon mode, clearly showing that the Raman redshift and hence the strain of MoS$_2$ is engineered in accordance to our device design (B). Center for Quantum Frontiers of Research \& Technology (QFort) authorizes the use of the logo which is shown in (B) and (E).}
\end{figure*}

Here we report on the control of both the strain magnitude and distribution in 2D materials with a high degree of freedom and surface integrity. We achieve this by conforming 2D molybdenum disulfide (MoS$_2$) -- one of the group-VI transition metal dichalcogenide (TMDC) semiconductors that are ideal for future electronic and optoelectronic devices due to many unprecedented properties from the combination of their 2D nature and spin-orbit coupling\cite{Lee_NM2017,Zeng_NT2012} -- to hexagonal boron nitride (hBN) substrates with lithographically-defined surface nanostructures. It is important to stress that using hBN for substrate engineering is significant, because the hBN is essential for realizing the full potential of 2D material-based devices and technologies as it effectively helps minimize the extrinsic disorder and dielectric loss\cite{Dean_NT2010, Rhodes_NM2019}, which plays a vital role in device technology, e.g. in the advent of next-generation transistors\cite{Lee_Science2018,Chen_Nature2020,Ma_Nature2022} and quantum logic\cite{Wang_NM2022}. Hence, the surface flatness and integrity of hBN must be retained while fabricating any desirable nanostructures upon it, for which we present the best surface flatness yet reported for any hBN etching. The ability to perform lithography and etch of hBN with atomic-level flatness enables extremely smooth arbitrary 3D nanoarchitecture to be created, bringing new possibilities for strain engineering when 2D materials are placed on the etched nanostructures.

We use micro-Raman and photoluminescence spectroscopy to demonstrate the on-chip controllability of strain within a single 2D flake, showing the ability to write and draw any desirable color image for optoelectronics. More strikingly, we report the observation of prominent interlayer excitons with surprisingly large oscillator strength in our high-quality multilayer strain-engineered MoS$_2$ devices at room temperature. Although the interlayer excitons---the quasiparticles that consist of strongly bound electrons and holes residing in separate layers---have been observed in many systems\cite{Gong_NM2014,Rivera_NC2015,Rivera_S2016,Rivera_NT2018,Deilmann_NL2018,Gerber_PRB_2019,Wang_N2019,Tran_N2019,Jauregui_S2019,Jin_N2019,Ciarrocchi_NP2019,Li_NM2020,Gu_NP2022,Niehues_Nanoscale2019,Paradisanos_NC2020,Yuan_NM2020,Barre_Science22,Xia_NP2021,Sun_Nature2022,Peimyoo_NT2021}, their strengths are normally much weaker compared to their intralayer counterparts. Hence, there is a considerable interest in finding ways to enlarge their strength and modulate their energy, e.g. by varying the twist angle between the constituent layers\cite{Paradisanos_NC2020,Yuan_NM2020,Barre_Science22}, engineering hydrostatic pressure\cite{Xia_NP2021}, or reducing dielectric screening through free-standing structures\cite{Sun_Nature2022}. The out-of-plane high-quality strain engineering reported here is shown to increase the strength ratio between the interlayer and intralayer excitons in MoS$_2$ from $\sim10~\%$ to more than $40~\%$, four times greater than in unstrained regions, at room temperature. Our strain engineering method provides a viable alternative to modulate both the energy and oscillator strength of interlayer excitions and could pave a way for promising excitonic devices and integrated circuits\cite{Mueller_npj18,Jiang_light21,Regan_nrm22}.

Figure~1A illustrates the fabrication process of our strain engineered MoS$_2$ optoelectronic devices (see Materials and Methods for fabrication details). A large flake of hBN is first prepared by mechanical exfoliation, followed by electron-beam lithography and plasma etching to create any desired surface nanostructure and topography. For example, we can imprint the letter `N' that consists with many 100-nm-wide trenches to form a corrugated surface. Creating nanostructures via etch on hBN is distinct from that on conventional substrates such as Si and SiO$_2$ since the hBN is a chemically inert 2D material without any dangling bounds. The extremely weak interlayer bounding (by van der Waals forces) with respect to the strong intralayer covalent bounding allows us develop etching techniques to engrave whatever desired geometric pattern and image on the hBN surface with atomic-layer precision and excellent selectivity, while retaining the surface cleanness and integrity. The surface roughness of our etched hBN on SiO$_2$ substrate is around $1$~{\AA} (Supplementary Fig.~S1), which is, to the best of our knowledge, the lowest value yet reported for any material etching (and about one order of magnitude smaller than earlier reported hBN etching\cite{Park_Nanoscale2018}). This therefore provides a unique playground to strain engineer high-quality 2D materials that are dry transferred onto it, e.g. MoS$_2$ in this work, and reveal the science of straintronics without being disturbed and hampered by extrinsic disorder.

The optical microscopy image of one of our strain engineered MoS$_2$/hBN devices is shown in Fig.~1B, clearly displaying our graphically designed pattern for strain engineering. Atomic force microscopy (AFM) is further employed to certify that the transferred monolayer MoS$_2$ membrane indeed conforms to the lithographically-defined nanostructures in the hBN substrate (Fig.~1C), as well as to examine the surface integrity and cleanness of the device (Fig.~1D). The surface roughness of MoS$_2$ film on the unetched and etched hBN regions are approximately $80$~pm and $120$~pm, respectively; both values are much smaller than that of the underlying SiO$_2$ substrate ($\sim 350$~pm). This approach enables us to create significantly smoother 2D material surfaces and interfaces than prior strain engineering techniques, essential for realizing high-quality 2D strained electronics and optoelectronics.

The most direct and straightforward technique to characterize strain in 2D materials is Raman spectroscopy. Figure~2F maps the band shifts of the $A_{1g}$ Raman-active mode observed in our strain-engineered monolayer MoS$_2$/hBN device, using an excitation wavelength of $532$~nm (see Methods). A redshift in Raman peaks only occurs in the regions with etched nanostructures (hereafter referred to as strained regions), thereby demonstrating that strain is induced in MoS$_2$ in accordance to our design.

\begin{figure}
\begin{center}
\includegraphics[width=1\columnwidth]{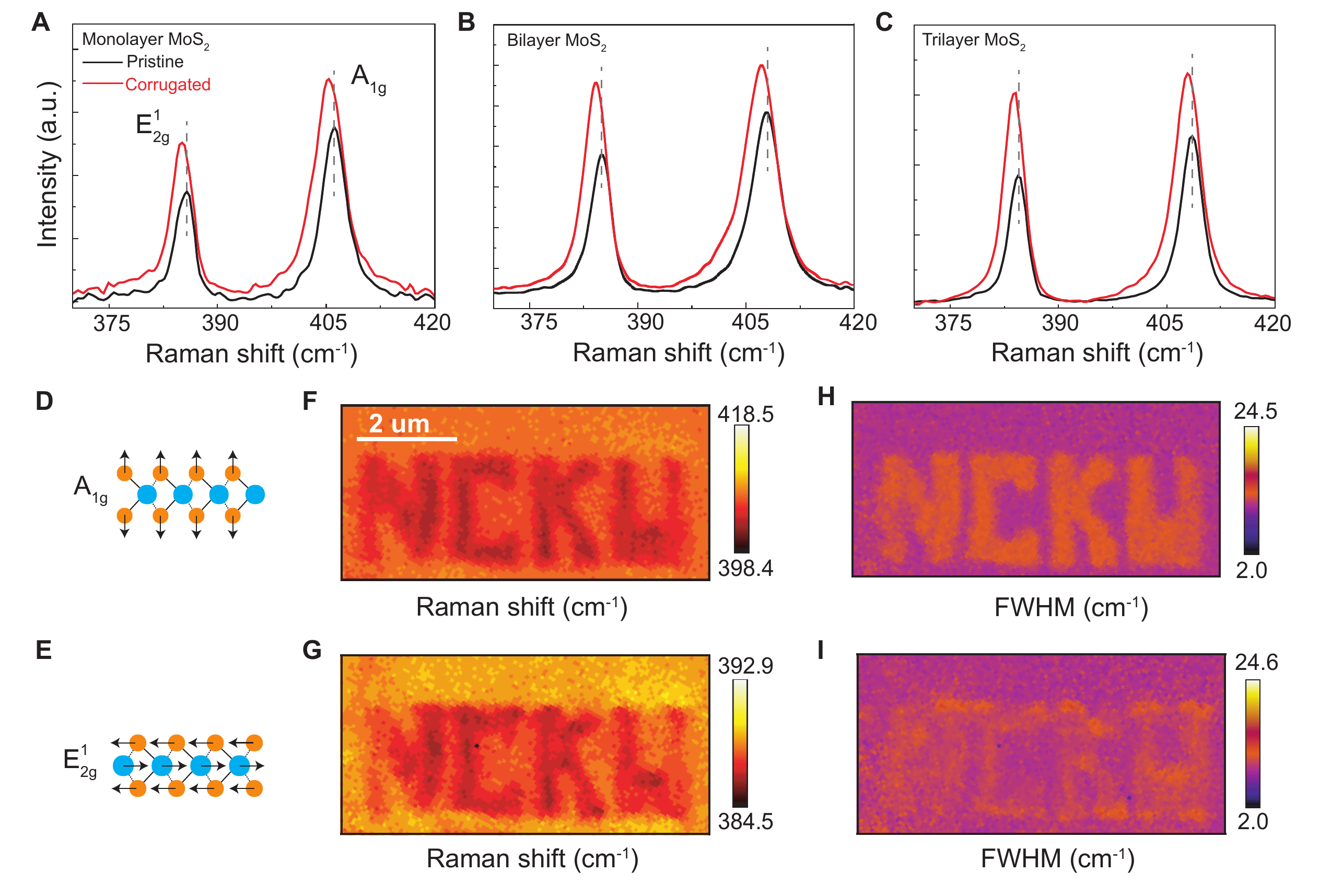}
\end{center}
\caption{\textbf{Raman spectra of strain engineered MoS$_2$/hBN devices.} (\textbf{A}) to (\textbf{C}) Comparison of pristine (black line) and strained (red line) MoS$_2$ Raman spectra with monolayer (A), bilayer (B) and trilayer (C). Grey dashed lines are located at peak maxima of pristine MoS$_2$ to illustrate the difference in peak position between pristine and strained MoS$_2$. (\textbf{D} and \textbf{E}) Schematic illustrations of the Raman mode with the out-of-plane (A$_{1g}$) phonon vibration and the in-plane (E$^1_{2g}$) phonon vibration, respectively. (\textbf{F} and \textbf{G}) Spatial maps of Raman spectroscopy peak positions for the strain engineered MoS$_2$ device corresponding to A$_{1g}$ and E$^1_{2g}$ Raman modes, respectively. (\textbf{H} and \textbf{I}) Same as (F) and (G) but mapping the full width at half maximum (FWHM) of the Raman peaks.}
\end{figure}

We now systematically examine the strain in such devices. Figures~2 (A to C) respectively show Raman spectra measured in mono-, bi- and tri-layer MoS$_2$ flakes, consisting of flat (unstrained) and corrugated (strain-engineered) regions within the same flake, to directly compare the impact of strain. The Raman peaks that are prominently observed and used for characterizing strain in TMDCs arise from the first-order in-plane $E^1_{2g}$ and out-of-plane $A_{1g}$ phonon modes (see Figs.~2, D and E for their corresponding atomic displacements). The phonon frequencies for these two modes in the flat unstrained regions (black traces) are $\sim 385$~cm$^{-1}$ ($E^1_{2g}$) and $\sim 408$~cm$^{-1}$ ($A_{1g}$), consistent with previous results for thin-layer pristine MoS$_2$\cite{Rice_PRB13,Conley_NL2013}. For spectra in the strain-engineered regions (red traces), redshifts of both the $E^1_{2g}$ and $A_{1g}$ modes are clearly observed, indicating that an uniaxial tensile strain is indeed introduced. The spatial maps of their peak positions further assert that the local distribution of strain can be well manipulated (Figs.~2, F and~G). Moreover, a peak broadening also occurs in addition to the redshift for both Raman modes. To have a better view of the peak broadening, we map the full width at half maximum (FWHM) for $E^1_{2g}$ and $A_{1g}$ modes in Figs.~2 (H~and~I), respectively. The reason for this is because the strain created in such out-of-plane nanostructures is non-uniform\cite{Levy_S2010,Ho_NE2021,Li_PRL20,Hsu_SA20,Kang_NV2021}. In fact, it is significant to have non-uniform strain as this is the key to lead to a nonzero pseudo-magnetic field and complex interlayer coupling, yielding rich physics and promising applications especially in valleytronics.

\begin{figure}
\begin{center}
\includegraphics[width=1\columnwidth]{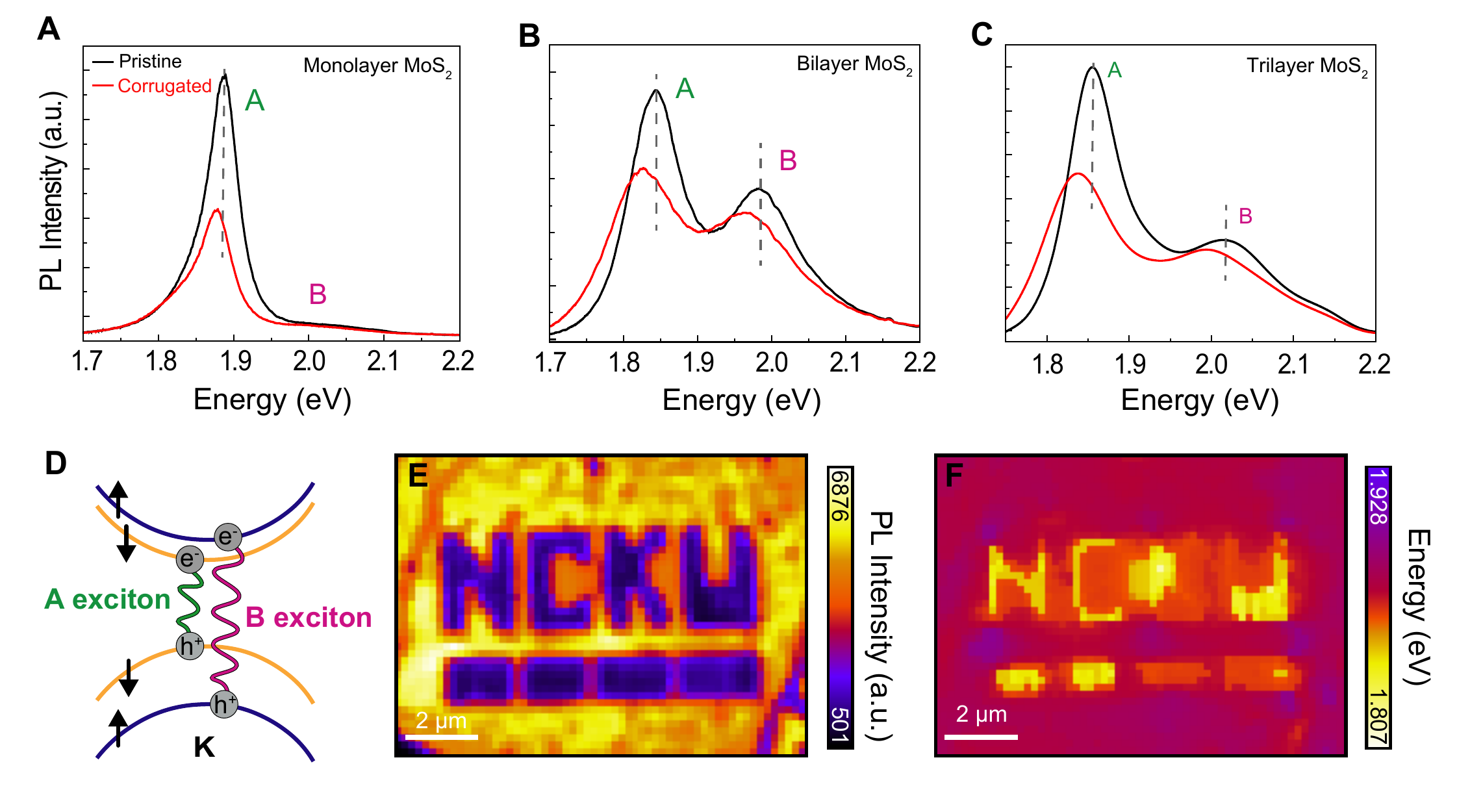}
\end{center}
\caption{\textbf{Photoluminescence spectra of strain engineered MoS$_2$/hBN devices.} (\textbf{A}) to (\textbf{C}) Comparison of pristine (black line) and strained (red line) MoS$_2$ PL spectra with monolayer (A), bilayer (B) and trilayer (C). Grey dashed lines are located at peak maxima of pristine MoS$_2$ to show the difference of peak position between pristine and strained MoS$_2$. (\textbf{D}) Schematic band diagram of MoS$_2$ around the K point with their corresponding spin directions labeled in black arrows. Two intralayer excitons A and B are represented by green and magenta wavy lines, respectively. (\textbf{E} and \textbf{F}) Spatial maps of photoluminescence peak intensity (E) and energy (F) of A exciton in the strain engineered monolayer MoS$_2$ device.}
\end{figure}

A thin layer of MoS$_2$ and other group-VI TMDCs have recently attracted tremendous interest owing to their remarkably rich and versatile excitonic states / effects that dominate the optical and optoelectronic responses even at room temperatures, together with their excellent properties for optoelectronic applications such as direct band gaps in the infrared and visible spectral range, strong light-matter interactions, and large spin-orbit couplings\cite{Mak_PRL2010,Splendiani_NL2010}. The unique combination of these properties is an exciting avenue for future exciton-based optoelectronics and spin/valleytronics, and it will be interesting to see whether strain engineering can open up more opportunities. To investigate the impact of strain on the electronic band structure and optical properties of MoS$_2$, we conduct micro-photoluminescence measurements. Figures~3 (A to C) show the photoluminescence (PL) spectra of mono-, bi- and tri-layer MoS$_2$, respectively, in both unstrained and strained regions. Redshifts are observed for both the A and B intralayer exciton emissions in the strained regions for all our MoS$_2$ devices, regardless of the thickness. The results are consistent with prior studies for uniaxial tensile strains introduced using other approaches\cite{Conley_NL2013,Castellanos_NL2013} and indicate that the strain indeed modifies band gap and optical response. The A and B exciton emissions originate from a lower- and higher-energy transition between the spin-orbit split conduction and valance bands, respectively, as shown in Fig.~3D. Note that in monolayer MoS$_2$ (Fig.~3A) the negligibly small emission of the B exciton indicates that our strain-engineering scheme of using the atomic-scale engraved hBN substrate produces devices of high quality and low defect density\cite{McCreary_apl_material}. Furthermore, we spatially map the peak intensity (Fig.~3E) and energy (Fig.~3F) of the A exciton emission of the monolayer MoS$_2$ device. This demonstrates that the excitonic states and optical propoerties of 2D TMDCs can be well manipulated in accordance to any graphic design of strain.

\begin{figure}
\begin{center}
\includegraphics[width=1\columnwidth]{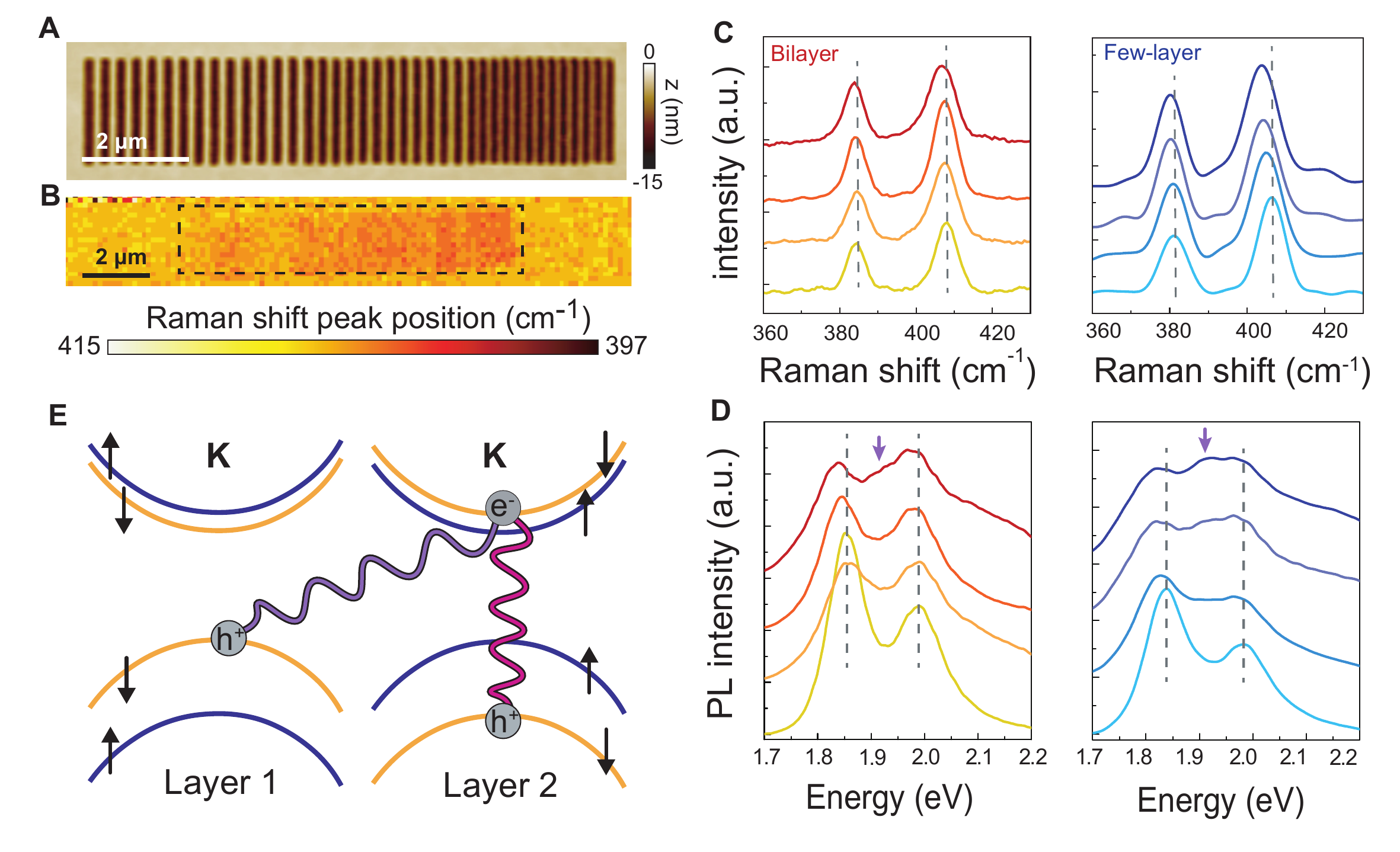}
\end{center}
\caption{\textbf{On-chip strain tunability.}
(\textbf{A}) AFM profile of the strain-tunable MoS$_2$ device with varying periodicity of trenches. (\textbf{B}) Spatial map of the Raman peak position for the A$_{1g}$ mode of the MoS$_2$ device shown in (A). The magnitude of the Raman redshift, a proxy to evaluate the magnitude of strain, increases with increasing the periodicity from left to right. (\textbf{C}) Raman spectra of strain engineered MoS$_2$ devices, showing the evolution of peak position of the E$^1_{2g}$ mode and A$_{1g}$ mode for different corrugation periodicities in bi- (left) and few-layer (right) MoS$_2$. The Raman vibration modes of the pristine MoS$_2$ are marked with the vertical grey dashed lines. (\textbf{D}) Photoluminescence spectra measured in the same strain engineered regions and devices corresponding to (C). The purple arrows indicate the emergence of an additional PL peak between the A and B exciton, attributed to the interlayer exciton, when the strain introduced to the system is relatively large. The traces in (C) and (D) are offset vertically for clarity. (\textbf{E}) Schematic illustration of interlayer exciton in bilayer MoS$_2$, which involves an electron localized in one layer and a hybridized hole state delocalized over two layers.}
\end{figure}

So far we have demonstrated control of the strain distribution, we now move on to show how the magnitude of strain can also be well engineered in our systems. This can be simply achieved, for example, by etching trenches with different dimensions and/or periodicities in the hBN substrate. Figure~4A shows an example AFM image of our device designs; in this particular design, the width and depth of the trenches are fixed while the spacing between them is varied from $130$ to $40$~nm (from left to right). A spatial map of the Raman spectroscopy for the $A_{1g}$ phonon mode in Fig.~4B shows that the magnitude of redshift, a proxy to evaluate the magnitude of strain, increases with decreasing the spacing between trenches. Figure~4C shows the Raman spectra at several different strain-controlled regions to give a complete comparison. Both phonon modes gradually shift to a lower energy following our strain design in all devices. These results demonstrate that the strain magnitude and distribution can both be well engineered to any designs even within a single 2D material flake. To quantify the strain magnitude, we follow the relation between the Raman shift and the uniaxial in-plane strain\cite{Rice_PRB13} and estimate that the strain ranges from $0.07~\%$ to $\sim 0.75~\%$ (Supplementary Section~2). Note that this may only give a very rough estimate since the strain in our system is different from the in-plane strain on a flexible substrate where this relation is applicable. Nevertheless, this clearly shows the tunability of the strain.

Figure~4D shows the PL spectra of both the bilayer and few-layer MoS$_2$ strained devices as the strain magnitude is gradually increased. The redshifts of A and B excitons are both observed to increase with increasing strain magnitude. This demonstrates that the frequency/wavelength of the exciton emissions can be manipulated within a single 2D flake using this strain engineering method, thereby opening opportunities for tuning colour in 2D TMDC optoelectronic devices. More intriguingly, we observe the emergence of an additional spectral feature at $\sim1.913$~eV (marked by arrows in Fig.~4D) when a relatively large strain is introduced in multi-layer MoS$_2$. Note that it remains absent in monolayer MoS$_2$ (Supplementary Fig.~S3). This additional spectral feature is attributed to the formation of interlayer excitons, which are associated with electrons and holes in different MoS$_2$ layers as depicted in Fig.~4E. The interlayer excitons have been observed in several (twisted) TMDC heterostructures\cite{Gong_NM2014,Rivera_NC2015,Rivera_S2016,Rivera_NT2018,Wang_N2019,Tran_N2019,Jauregui_S2019,Jin_N2019,Ciarrocchi_NP2019,Li_NM2020,Gu_NP2022,Barre_Science22} and homobilayer 2H-MoS$_2$\cite{Deilmann_NL2018,Gerber_PRB_2019,Niehues_Nanoscale2019,Paradisanos_NC2020,Peimyoo_NT2021}, continually sparking interest in novel exciton physics and technological applications because of their superior characteristics in several aspects\cite{Mueller_npj18,Jiang_light21,Regan_nrm22}, e.g. ultrafast formation and long population lifetime. The formation of interlayer excitons are known to be sensitive to the interlayer coupling and hence our non-uniform out-of-plane strain engineering, which has been shown to be able to vary the interlayer coupling and hopping mechanisms\cite{Ho_NE2021}, can be an alternative and viable means to modulate the interlayer excitons.

\begin{figure}
\begin{center}
\includegraphics[width=1\columnwidth]{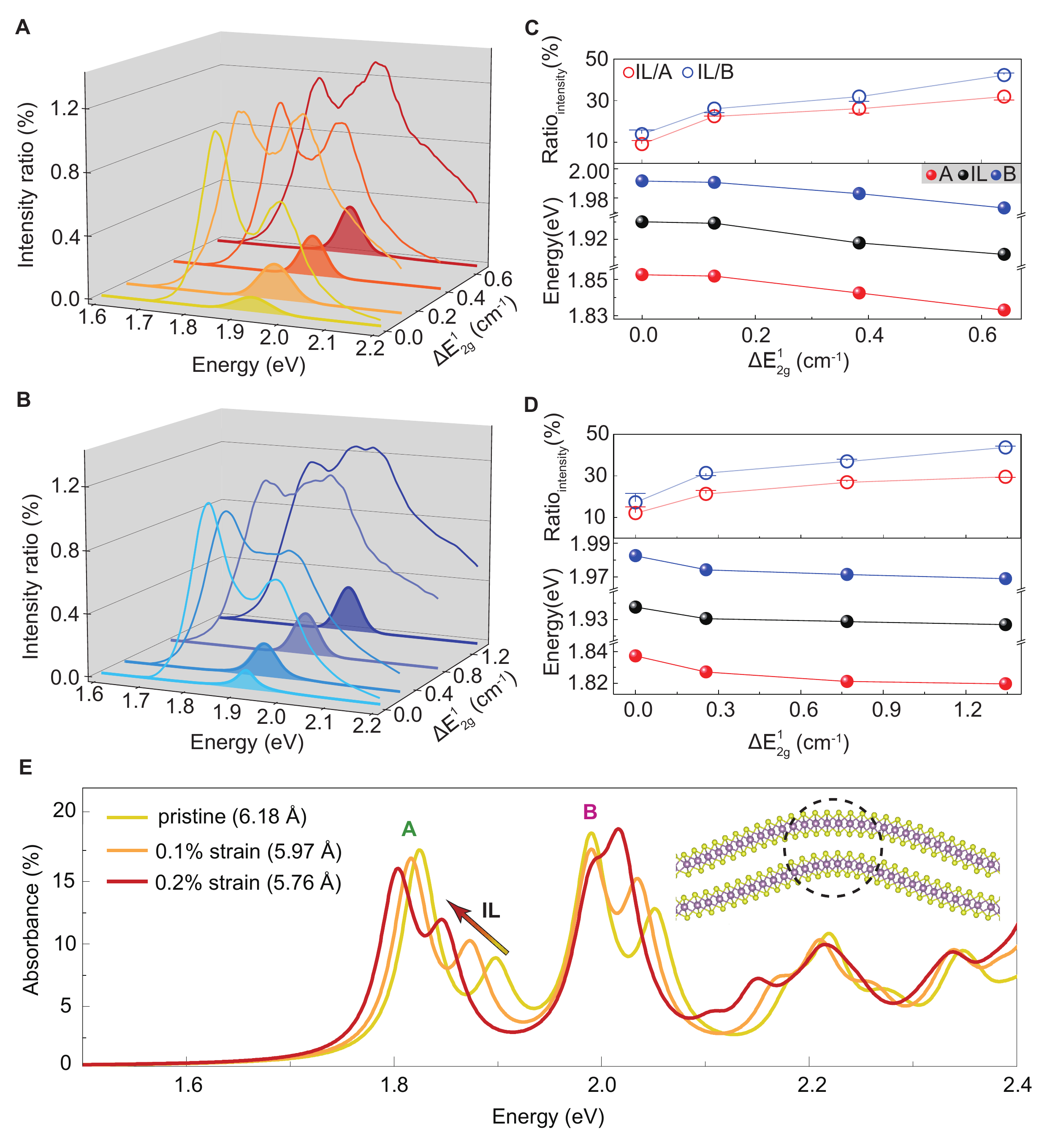}
\end{center}
\caption{\textbf{Strain tunable interlayer excitons.}
(\textbf{A} and \textbf{B}) Evolution of the decomposed PL spectra of the interlayer excitons as a function of the strain magnitude as quantified using the $\Delta E^1_{2g}$, in bilayer (A) and few-layer (B) MoS$_2$. The complete PL spectra are also plotted for comparison. All spectra are normalized by the decomposed A exciton intensity. (\textbf{C} and \textbf{D}) Upper panels show the intensity ratio between the interlayer excitons and the A (red) and B (blue) excitons at different strain magnitude quantified by $\Delta E^1_{2g}$ for bilayer (C) and few-layer MoS$_2$ (D). Lower panels show the PL peak energy of the A excitons (red), interlayer excitons (black), and B excitons (blue) as a function $\Delta E^1_{2g}$ in bilayer (C) and few-layer (D) MoS$_2$. The error bars represent the maximum variation in the intensity ratio when adjusting the energy of the interlayer exciton within a range of ±1.5 meV, accounting for possible uncertainty. (\textbf{E}) The optical spectrum of bilayer MoS$_2$ calculated using different strain magnitudes and interlayer distances, the values of which are estimated from the middle part of a hump structure as illustrated in the inset.}
\end{figure}

To further analyze the impact of strain on excitons, we deduce the interlayer excitonic components using Voigt fits of the PL spectra at different strain magnitudes in Figs.~5 (A and B) for the bi- and few-layer MoS$_2$, respectively (Supplementary Section~4); data are obtained from the same MoS$_2$ flakes as in Fig.~4D for a direct comparison. It is shown that the optical transitions from the interlayer excitons becomes more pronounced with increasing strain magnitude. We further compare PL peak intensity of the interlayer excitons with those of the intralayer A and B excitons as a function of the Raman shifts, which quantify the strain magnitude, in the upper panels of Figures~5 (C and D). The PL intensities for the interlayer exciton transitions are around $10 \%$ of the intralayer transitions in the flat unstrained regions. These ratios increase to more than $40 \%$ with increasing the strain magnitude, four times greater than those in unstrained regions. In addition to the enhancement of the interlayer excitons, their PL peak positions have also been observed to shift to a lower energy as the strain magnitude increases, consistent with the intralayer excitons (Figs.~5, C and D, lower panels). This indicates that both the intra- and interlayer bandgaps can be modulated by our strain engineering. It is important to stress that the out-of-plane non-uniform strain engineering reported here is distinctly different from the in-plane uniform strain created using a flexible substrate. This is because the in-plane uniform strain (lattice deformation) hardly modulates the interlayer coupling and thus the interlayer excitions, for which only redshifts in the exciton resonance energy are expected\cite{Niehues_Nanoscale2019}. 

To understand the possible mechanisms for the enhancement of interlayer excitons in our strained systems, we conduct density-functional theory (DFT) calculations on a uniaxially strained bi-layer MoS$_2$ on a monolayer hBN hump (Supplementary Section~$5$), which simulates the corner structure of corrugated trench. A side view of the setup after structure optimization is shown in the inset of Fig.~5E. We find that the interlayer distance around the corner edge (as marked in dashed circle) shrinks largely to $5.76$~{\AA} compared to its unstrained value of $6.18$~{\AA}. Due to the funnel effects\cite{Castellanos_NL2013}, the PL spectrum should be dominated by emission from these local regions. The optical spectrum calculated using such a horizontally stretched but vertically compressed bilayer unit cell MoS$_2$ is shown in Fig.~5E, which qualitatively agrees with the experiments. We acknowledge that this is an extremely simplified structure compared to the experiment. It is also known the out-of-plane non-uniform strain induces enormous pseudo-magnetic fields\cite{Guinea_2010,Levy_S2010} -- which dramatically change the electronic structure and properties -- and can also modulate the interlayer coupling in a nontrivial way\cite{Ho_NE2021}. Hence, a more sophisticated calculation considering large scale structure and other strain induced effects could be an interesting future study. Note that the change of the interlayer coupling is also evidenced by the observation of the unusually large Raman shifts in the out-of-plane $A_{1g}$ phonon modes (Supplementary Section~2).

In summary, we report the advent of hBN etching with atomic flatness and excellent selectivity. This enables an arbitrary control of strain both in their spatial distribution and magnitude of high-quality TMDC semiconductors using a surface-engineered hBN substrate. Spatially resolved Raman and PL mappings and spectrum analyses demonstrate that such strain engineering can modify the phonon scatterings, electronic band structures, and thus the exciton resonances at each local spatial region within a single 2D device. In addition, we observe a large enhancement of the interlayer exciton transitions as the strain increases due to the strain-induced interlayer coupling modification. The high quality and controllability of our strain engineering method, together with the ability to substantially modify the strengths and energies of interlayer excitions at room temperature, will open up more opportunities in 2D excitonics, optoelectronics, and straintronics.

\paragraph{Acknowledgements}
We thank L. W. Smith, J. I.-J. Wang and C.-K. Yong for helpful discussions. This work is supported by the National Science and Technology Council in Taiwan (grant numbers 110-2628-M-006-004, 110-2124-M-006-007-MY3, 110-2628-M-006-002-MY3, 110-2112-M-001-018-MY3), and the Higher Education Sprout Project, Ministry of Education to the Headquarters of University Advancement at the National Cheng Kung University (NCKU). K.W. and T.T. acknowledge support from JSPS KAKENHI (Grant Numbers 19H05790, 20H00354 and 21H05233).


\begin{thebibliography}{99}

\bibitem{Ni_2008} Z. H. Ni, T. Yu, Y. H. Lu, Y. Y. Wang, Y. P. Feng, Z. X. Shen, Uniaxial strain on graphene: Raman spectroscopy study and band-gap opening. \textit{ACS Nano}\textbf{2}, 2301-2305 (2008).
\bibitem{Guinea_2010} F. Guinea, M. Katsnelson, A. Geim, Energy gaps and a zero-field quantum Hall effect in graphene by strain engineering.\textit{Nature Physics} \textbf{6}, 30–33 (2010).
\bibitem{Si_2013} C. Si, Z. Liu, W. Duan, F. Liu, First-principles calculations on the effect of doping and biaxial tensile strain on electron-phonon coupling in graphene. \textit{Phys. Rev. Lett.} \textbf{111}, 196802 (2013).
\bibitem{Ho_NE2021} S.-C. Ho, C.-H. Chang, Y.-C. Hsieh, S.-T. Lo, B. Huang, T.-H.-Y. Vu, C. Ortix, T.-M. Chen, Hall effects in artificially corrugated bilayer graphene without breaking time-reversal symmetry. \textit{Nature Electronics} \textbf{4}, 116-125 (2021).
\bibitem{Dai_AM2019} Z. Dai, L. Liu, Z. Zhang, Strain engineering of 2D materials: issues and opportunities at the interface. \textit{Advanced Materials} \textbf{31}, 1805417 (2019).

\bibitem{Rice_PRB13} C. Rice, R. Young, R. Zan, U. Bangert, D. Wolverson, T. Georgiou, R. Jalil, K. Novoselov, Raman-scattering measurements and first-principles calculations of strain-induced phonon shifts in monolayer MoS$_2$. \textit{Physical Review B} \textbf{87}, 081307 (2013).
\bibitem{Conley_NL2013} H. J. Conley, B. Wang, J. I. Ziegler, R. F. Haglund Jr, S. T. Pantelides, K. I. Bolotin, Bandgap engineering of strained monolayer and bilayer MoS$_2$. \textit{Nano Letters} \textbf{13}, 3626-3630 (2013).
\bibitem{Castellanos_NL2013} A. Castellanos-Gomez, R. Roldán, E. Cappelluti, M. Buscema, F. Guinea, H. S. van der Zant, G. A. Steele, Local strain engineering in atomically thin MoS$_2$. \textit{Nano Letters} \textbf{13}, 5361-5366 (2013).

\bibitem{Levy_S2010} N. Levy, S. Burke, K. Meaker, M. Panlasigui, A. Zettl, F. Guinea, A. C. Neto, M. F. Crommie, Strain-induced pseudo–magnetic fields greater than 300 tesla in graphene nanobubbles. \textit{Science} \textbf{329}, 544-547 (2010).

\bibitem{Li_PRL20} S.-Y. Li, Y. Su, Y.-N. Ren, L. He, Valley Polarization and Inversion in Strained Graphene via Pseudo-Landau Levels, Valley Splitting of Real Landau Levels, and Confined States. \textit{Phys. Rev. Lett.} \textbf{124}, 106802 (2020).


\bibitem{Hsu_SA20} C.-C. Hsu, M. Teague, J.-Q. Wang, N.-C. Yeh, Nanoscale strain engineering of giant pseudo-magnetic fields, valley polarization, and topological channels in graphene. \textit{Science Advances} \textbf{6}, eaat9488 (2020).
\bibitem{Kang_NV2021} D.-H. Kang, H. Sun, M. Luo, K. Lu, M. Chen, Y. Kim, Y. Jung, X. Gao, S. J. Parluhutan, J. Ge, Pseudo-magnetic field-induced slow carrier dynamics in periodically strained graphene. \textit{Nature Communications} \textbf{12}, 1-8 (2021).

\bibitem{Li_NC2015} H. Li, A. W. Contryman, X. Qian, S. M. Ardakani, Y. Gong, X. Wang, J. M. Weisse, C. H. Lee, J. Zhao, P. M. Ajayan, Optoelectronic crystal of artificial atoms in strain-textured molybdenum disulphide. \textit{Nature Communications} \textbf{6}, 1-7 (2015).
\bibitem{Li_NM2016} H. Li, C. Tsai, A. L. Koh, L. Cai, A. W. Contryman, A. H. Fragapane, J. Zhao, H. S. Han, H. C. Manoharan, F. Abild-Pedersen, Activating and optimizing MoS$_2$ basal planes for hydrogen evolution through the formation of strained sulphur vacancies. \textit{Nature Materials} \textbf{15}, 48-53 (2016).
\bibitem{Choi_NL2015} J. Choi, J. Mun, M. C. Wang, A. Ashraf, S.-W. Kang, S. Nam, Hierarchical, dual-scale structures of atomically thin MoS$_2$ for tunable wetting. \textit{Nano Letters} \textbf{17}, 1756-1761 (2017).


\bibitem{Zeng_NT2012} H. Zeng, J. Dai, W. Yao, D. Xiao, X. Cui, Valley polarization in MoS$_2$ monolayers by optical pumping. \textit{Nature Nanotechnology} \textbf{7}, 490-493 (2012).

\bibitem{Lee_NM2017} J. Lee, Z. Wang, H. Xie, K. F. Mak, J. Shan, Valley magnetoelectricity in single-layer MoS$_2$. \textit{Nature Materials} \textbf{16}, 887-891 (2017).
\bibitem{Dean_NT2010} C. R. Dean, A. F. Young, I. Meric, C. Lee, L. Wang, S. Sorgenfrei, K. Watanabe, T. Taniguchi, P. Kim, K. L. Shepard, Boron nitride substrates for high-quality graphene electronics. \textit{Nature Nanotechnology} \textbf{5}, 722-726 (2010).
\bibitem{Rhodes_NM2019} D. Rhodes, S. H. Chae, R. Ribeiro-Palau, J. Hone, Disorder in van der Waals heterostructures of 2D materials. \textit{Nature Materials} \textbf{18}, 541-549 (2019).
\bibitem{Lee_Science2018} J. S. Lee, S. H. Choi, S. J. Yun, Y. I. Kim, S. Boandoh, J.-H. Park, B. G. Shin, H. Ko, S. H. Lee, Y.-M. Kim, Wafer-scale single-crystal hexagonal boron nitride film via self-collimated grain formation. \textit{Science} \textbf{362}, 817-821 (2018). 
\bibitem{Chen_Nature2020} T.-A. Chen, C.-P. Chuu, C.-C. Tseng, C.-K. Wen, H.-S. P. Wong, S. Pan, R. Li, T.-A. Chao, W.-C. Chueh, Y. Zhang, Wafer-scale single-crystal hexagonal boron nitride monolayers on Cu (111). \textit{Nature} \textbf{579}, 219–223 (2020).
\bibitem{Ma_Nature2022} K. Y. Ma, L. Zhang, S. Jin, Y. Wang, S. I. Yoon, H. Hwang, J. Oh, D. S. Jeong, M. Wang, S. Chatterjee, Epitaxial single-crystal hexagonal boron nitride multilayers on Ni (111). \textit{Nature} \textbf{606}, 88-93 (2022).
\bibitem{Wang_NM2022} J. I. Wang, M. A. Yamoah, Q. Li, A. H. Karamlou, T. Dinh, B. Kannan, J. Braumüller, D. Kim, A. J. Melville, S. E. Muschinske, Hexagonal boron nitride as a low-loss dielectric for superconducting quantum circuits and qubits. \textit{Nature Materials} \textbf{21}, 398-403 (2022).
\bibitem{Rivera_NC2015} P. Rivera, J. R. Schaibley, A. M. Jones, J. S. Ross, S. Wu, G. Aivazian, P. Klement, K. Seyler, G. Clark, N. J. Ghimire, Observation of long-lived interlayer excitons in monolayer MoSe$_2$–WSe$_2$ heterostructures. \textit{Nature Communications} \textbf{6}, 1-6 (2015).
\bibitem{Rivera_S2016} P. Rivera, K. L. Seyler, H. Yu, J. R. Schaibley, J. Yan, D. G. Mandrus, W. Yao, X. Xu, Valley-polarized exciton dynamics in a 2D semiconductor heterostructure. \textit{Science} \textbf{351}, 688–691 (2016).
\bibitem{Tran_N2019} K. Tran, G. Moody, F. Wu, X. Lu, J. Choi, K. Kim, A. Rai, D. A. Sanchez, J. Quan, A. Singh, Evidence for moiré excitons in van der Waals heterostructures. \textit{Nature} \textbf{567}, 71-75 (2019).



\bibitem{Li_NM2020} W. Li, X. Lu, S. Dubey, L. Devenica, A. Srivastava, Dipolar interactions between localized interlayer excitons in van der Waals heterostructures. \textit{Nature Materials} \textbf{19}, 624-629 (2020).
\bibitem{Gong_NM2014} Y. Gong, J. Lin, X. Wang, G. Shi, S. Lei, Z. Lin, X. Zou, G. Ye, R. Vajtai, B. I. Yakobson, Vertical and in-plane heterostructures from WS$_2$/MoS$_2$ monolayers. \textit{Nature Materials} \textbf{13}, 1135-1142 (2014).
\bibitem{Rivera_NT2018} P. Rivera, H. Yu, K. L. Seyler, N. P. Wilson, W. Yao, X. Xu, Interlayer valley excitons in heterobilayers of transition metal dichalcogenides. \textit{Nature Nanotechnology} \textbf{13}, 1004-1015 (2018).
\bibitem{Jin_N2019} C. Jin, E. C. Regan, A. Yan, M. Iqbal Bakti Utama, D. Wang, S. Zhao, Y. Qin, S. Yang, Z. Zheng, S. Shi, Observation of moiré excitons in WSe$_2$/WS$_2$ heterostructure superlattices. \textit{Nature} \textbf{567}, 76-80 (2019).
\bibitem{Jauregui_S2019} L. A. Jauregui, A. Y. Joe, K. Pistunova, D. S. Wild, A. A. High, Y. Zhou, G. Scuri, K. De Greve, A. Sushko, C.-H. Yu, Electrical control of interlayer exciton dynamics in atomically thin heterostructures. \textit{Science} \textbf{366}, 870-875 (2019).
\bibitem{Ciarrocchi_NP2019} A. Ciarrocchi, D. Unuchek, A. Avsar, K. Watanabe, T. Taniguchi, A. Kis, Polarization switching and electrical control of interlayer excitons in two-dimensional van der Waals heterostructures. \textit{Nature photonics} \textbf{13}, 131-136 (2019).
\bibitem{Gu_NP2022} J. Gu, L. Ma, S. Liu, K. Watanabe, T. Taniguchi, J. C. Hone, J. Shan, K. F. Mak, Dipolar excitonic insulator in a moire lattice. \textit{Nature Physics} \textbf{18}, 395-400 (2022).





\bibitem{Wang_N2019}K. Tran, G. Moody, F. Wu, X. Lu, J. Choi, K. Kim, A. Rai, D. A. Sanchez, J. Quan, A. Singh, Evidence of high-temperature exciton condensation in two-dimensional atomic double layers. \textit{Nature} \textbf{574}, 76-80 (2019).

\bibitem{Deilmann_NL2018} T. Deilmann, K. S. Thygesen, Interlayer excitons with large optical amplitudes in layered van der Waals materials. \textit{Nano Letters} \textbf{18}, 2984-2989 (2018).
\bibitem{Gerber_PRB_2019} I. C. Gerber, E. Courtade, S. Shree, C. Robert, T. Taniguchi, K. Watanabe, A. Balocchi, P. Renucci, D. Lagarde, X. Marie, Interlayer excitons in bilayer MoS$_2$ with strong oscillator strength up to room temperature. \textit{Physical Review B} \textbf{99}, 035443 (2019).
\bibitem{Niehues_Nanoscale2019} I. Niehues, A. Blob, T. Stiehm, S. M. de Vasconcellos, R. Bratschitsch, Interlayer excitons in bilayer MoS$_2$ under uniaxial tensile strain. \textit{Nanoscale} \textbf{11}, 12788-12792 (2019).


\bibitem{Peimyoo_NT2021} N. Peimyoo, T. Deilmann, F. Withers, J. Escolar, D. Nutting, T. Taniguchi, K. Watanabe, A. Taghizadeh, M. F. Craciun, K. S. Thygesen, Electrical tuning of optically active interlayer excitons in bilayer MoS$_2$. \textit{Nature Nanotechnology} \textbf{16}, 888-893 (2021).


\bibitem{Paradisanos_NC2020} I. Paradisanos, S. Shree, A. George, N. Leisgang, C. Robert, K. Watanabe, T. Taniguchi, R. J. Warburton, A. Turchanin, X. Marie, Controlling interlayer excitons in MoS$_2$ layers grown by chemical vapor deposition. \textit{Nature Communications} \textbf{11}, 1-7 (2020).



\bibitem{Yuan_NM2020} L. Yuan, B. Zheng, J. Kunstmann, T. Brumme, A. B. Kuc, C. Ma, S. Deng, D. Blach, A. Pan, L. Huang, Twist-angle-dependent interlayer exciton diffusion in WS$_2$–WSe$_2$ heterobilayers. \textit{Nature Materials} \textbf{19}, 617-623 (2020).

\bibitem{Barre_Science22} E. Barr\'e, O. Karni, E. Liu, A. L. O'Beirne, X. Chen, H. B. Ribeiro, L. Yu, B. Kim, K. Watanabe, T. Taniguchi, K. Barmak, C. H. Lui, S. Refaely-Abramson, F. H. da Jornada, T. F. Heinz, Optical absorption of interlayer excitons in transition-metal dichalcogenide heterostructures. \textit{Science} \textbf{376}, 406-410 (2022).



\bibitem{Xia_NP2021} J. Xia, J. Yan, Z. Wang, Y. He, Y. Gong, W. Chen, T. C. Sum, Z. Liu, P. M. Ajayan, Z. Shen, Strong coupling and pressure engineering in WSe$_2$–MoSe$_2$ heterobilayers. \textit{Nature Physics} \textbf{17}, 92-98 (2021).

\bibitem{Sun_Nature2022} X. Sun, Y. Zhu, H. Qin, B. Liu, Y. Tang, T. Lü, S. Rahman, T. Yildirim, Y. Lu, Enhanced interactions of interlayer excitons in free-standing heterobilayers. \textit{Nature} \textbf{610}, 478–484 (2022).





\bibitem{Mueller_npj18} T. Mueller, E. Malic, Exciton physics and device application of two-dimensional transition metal dichalcogenide semiconductors. \textit{npj 2D Materials and Applications} \textbf{2}, 1-12 (2018).

\bibitem{Jiang_light21} Y. Jiang, S. Chen, W. Zheng, B. Zheng, A. Pan, Interlayer exciton formation, relaxation, and transport in TMD van der Waals heterostructures. \textit{Light: Science \& Applications} \textbf{10}, 1-29 (2021).

\bibitem{Regan_nrm22}  E. C. Regan, D. WAng, E. Y. Paik, Y. Zeng, L. Zhang, J. Zhu, A. H. MacDonald, H. Deng, F. Wang, Emerging exciton physics in transition metal dichalcogenide heterobilayers \textit{Nat Rev Mater} \textbf{7}, 778-195 (2022).


\bibitem{Park_Nanoscale2018} H. Park, G. H. Shin, K. J. Lee, S.-Y. Choi, Atomic-scale etching of hexagonal boron nitride for device integration based on two-dimensional materials. \textit{Nanoscale} \textbf{10}, 15205-15212 (2018).

\bibitem{Mak_PRL2010} K. F. Mak, C. Lee, J. Hone, J. Shan, T. F. Heinz, Atomically thin MoS$_2$: a new direct-gap semiconductor. \textit{Phys. Rev. Lett.} \textbf{105}, 136805 (2010).
\bibitem{Splendiani_NL2010} A. Splendiani, L. Sun, Y. Zhang, T. Li, J. Kim, C.-Y. Chim, G. Galli, F. Wang, Emerging photoluminescence in monolayer MoS$_2$. \textit{Nano Letters} \textbf{10}, 1271-1275 (2010).
\bibitem{McCreary_apl_material} K. M. McCreary, A. T. Hanbicki, S. V. Sivaram, B. T. Jonker, A-and B-exciton photoluminescence intensity ratio as a measure of sample quality for transition metal dichalcogenide monolayers. \textit{APL Materials} \textbf{6}, 111106 (2018).







\end{thebibliography}
\end{document}